\documentclass[aps,prd,preprint,nofootinbib]{revtex4}
\usepackage{bbm}
\usepackage{amsfonts}
\usepackage{epsfig}
\usepackage{graphicx}
\newcommand{\bqa}{\begin{eqnarray}}
\newcommand{\eqa}{\end{eqnarray}}
\newcommand{ \slashchar }[1]{\setbox0=\hbox{$#1$}   
   \dimen0=\wd0                                     
   \setbox1=\hbox{/} \dimen1=\wd1                   
   \ifdim\dimen0>\dimen1                            
      \rlap{\hbox to \dimen0{\hfil/\hfil}}          
      #1                                            
   \else                                            
      \rlap{\hbox to \dimen1{\hfil$#1$\hfil}}       
      /                                             
   \fi}                                             %
\begin{document}
\title{Pseudoscalar Quarkonium Exclusive Decays to Vector Meson Pair\\[7mm]}

\author{Peng Sun$^{1}$, Gang Hao$^{1}$, Cong-Feng Qiao$^{1,2}
\footnote{corresponding author}$ } \affiliation{$^{1}$College of
Physical Sciences, Graduate University of Chinese Academy of
Sciences \\ YuQuan Road 19A, Beijing 100049, China}
\affiliation{$^{2}$Theoretical Physics Center for Science Facilities
(TPCSF), CAS\\ YuQuan Road 19B, Beijing 100049, China}

\author{~\vspace{0.9cm}}


\begin{abstract}
\vspace{3mm} The pseudoscalar quarkonia exclusive decays to light
mesons still poses a challenge to the theoretical understanding of
quarkonium properties in decay. In this work, we evaluate the
processes of pseudoscalar heavy quarkonium decays into vector meson
pairs, especially the helicity suppressed processes of
$\eta_b\rightarrow J/\psi J/\psi$ and $\eta_c\rightarrow VV$. In the
frame of NRQCD, the branching fraction of $Br[\eta_b\rightarrow
J/\psi J/\psi]$ are evaluated at the next-to-leading order of
perturbative QCD; and within the light-cone distribution formalism,
we calculate also the higher twist effects in these processes.
Numerical results show that the higher twist terms contribute more
than what from the NLO QCD corrections in the process of
$\eta_b\rightarrow J/\psi J/\psi$. It is found that the experimental
results on $\eta_c\rightarrow VV$ are hard to be understood by
merely the quark model and perturbative QCD calculation.

\vspace {7mm} \noindent {PACS number(s): 12.38.Bx, 13.25.Gv,
14.40.Pq }

\end{abstract}

\maketitle \section{Introduction}

In high energy physics, heavy quarkonium study is one of the most
interesting fields and it plays an important role in the
understanding of the configurations of hadrons and the
nonperturbative behavior of strong interaction. On one hand, the
heavy quark masses enable the perturbative QCD(pQCD) calculation for
quarkonium production and decay possible. On the other hand, due to
the non-relativistic nature of heavy quarkonium, one may investigate
their properties through a more transparent way, i.e. the effective
theory of non-relativistic QCD(NRQCD) \cite{NRQCD}.

It is well known that the S-wave spin-triplet heavy quarkonium
states, the $J/\psi$ and $\Upsilon$, can be produced directly in
$e^+e^-$ annihilation, and be measured via lepton pair decay mode
distinctively. These characters lead to rich experimental data and
deep investigations on them. While for their spin-singlet partners,
the $\eta_b$ and $\eta_c$, things are not that easy. At present,
people know relatively much less about their properties, especially
for $\eta_b$. For $\eta_c$, though there have been many measurements
in experiment on its various decay modes, puzzles remain in
confronting theoretical explanations to the experimental data, such
as in $\eta_c$ decay to vector meson pair \cite{etacvv,reviews}. For
$\eta_b$ study, there have many theoretical scenarios been put
forward \cite{s11,s12,b12,b13,b14,b15,b16}, and several experiments
been conducted \cite{ep1,ep2,ep3,ep4,ep5}. However, it was fixed
only in very recently by BaBar collaboration through $\Upsilon(3S)
\rightarrow \eta_b + \gamma$ process \cite{b11} and later on
confirmed by CLEO-c experiment \cite{cleoc}. About $\eta_b$ so far
we merely know the mass, its other properties are remaining unclear
and waiting for further investigations. It is worth noting that both
Babar and CLEO-c measurements are indirect ones. For further study
on $\eta_b$ physics, direct measurements on its decay products are
necessary.


For the direct $\eta_b$ detection, Braaten {\it et al.} suggested to
measure its exclusive decay products, the $J/\psi$ pair in $\eta_b$
decays \cite{b15}. In comparison with the experimental result for
$\eta_c\rightarrow \phi\phi$ and by some scaling assumptions, they
estimated the branching ratio of $\eta_b \rightarrow J/\psi J/\psi$
mode to be $7\times10^{-4\pm1}$, which hence is promising to be
observed in the Fermilab Tevatron Run II experiment. So far there
has been no conclusive report from the experiment yet, and the
theoretical estimation was questioned by Maltoni and Polosa
\cite{maltoni}. In the expectation of helicity conservation rules
\cite{HHC}, the leading order calculation in the nonrelativistic
limit gives null result. The calculations on next-to-leading order
QCD \cite{b12} and relativistic corrections \cite{b13} both yield
the branching ratios to be about $10^{-8}$. Recently, Braguta {\it
et al}. reevaluate the $\eta_b \rightarrow J/\psi J/\psi$ process in
the light cone formalism and find that after including the
next-to-leading twist contribution the branching fraction can be as
large as $(6.2\pm3.5)\times10^{-7}$ \cite{b14}. The authors of Ref.
\cite{b14} claim that the result in \cite{b13} dose not agree with
theirs. The form factor obtained in Ref.\cite{b13} contains double
logarithms, whereas they are absent in \cite{b14}. To carry on an
independent calculation of the $\eta_b \rightarrow J/\psi J/\psi$
process is therefore one of the aims of this work.

Similarly, the processes $\eta_c\rightarrow VV$ are also governed by
the helicity selection rules, but experiment gives extremely large
results \cite{etacvv}, which stands as a long term puzzle existing
in the charmonium physics. The higher order radiative corrections
give this issue no help, since they are all suppressed by the light
quark masses, Although beyond the scope of pQCD, some
nonperturbative models have been put forward and considered to be
solutions to the problem, such as the intermediate meson exchange
model \cite{zhaoqiang} and the charmonium light Fock component
admixture model \cite{feldman}, to further investigate it in pQCD is
still necessary.

The rest of this paper is organized as follows: in section II, we
calculate the branching ratio of the process $\eta_b \rightarrow
J/\psi J/\psi$ at one-loop level; in section III, we evaluate the
higher twist effects in processes $\eta_b \rightarrow J/\psi J/\psi$
and $\eta_c \rightarrow VV$; in section IV, summary and conclusions
are presented.

\section{NLO QCD result for $\eta_b \rightarrow J/\psi + J/\psi$ process}

In this section, we calculate the branching ratio of the process
$\eta_b \rightarrow J/\psi + J/\psi$ in the framework of NRQCD at
one-loop level and in non-relativistic limit. Hence, the relations
$p_b=p_{\bar b}=\frac{P_{\eta_b}}{2}$,
$p_{c_1}=p_{\bar{c}_1}=\frac{P_{J/\psi_1}}{2}$ and
$p_{c_2}=p_{\bar{c}_2}=\frac{P_{J/\psi_2}}{2}$ are adopted. The
bi-spinor operators are projected to states with the same quantum
numbers as $\eta_b$ and $\eta_c$, respectively, like
\bqa
v(p_{\bar b})\,\overline{u}(p_b)& \longrightarrow& {-1\over 2
\sqrt{2}\; m_b}\;(\frac{\not\! P_{\eta_b}}{2} + m_b)\;
\gamma_5\,(\frac{\not\! P_{\eta_b}}{2} - m_b)\otimes \left( {{\bf
1}_c\over \sqrt{N_c}}\right), \label{hc:projector1} \eqa
and
\bqa
v(p_{\bar c})\,\overline{u}(p_c)& \longrightarrow& {-1\over 2
\sqrt{2}\; m_c}\;(\frac{\not\! P_{J/\psi}}{2} - m_c)\; \not\!
\varepsilon^*\,(\frac{\not\! P_{J/\psi}}{2} + m_c)\otimes \left(
{{\bf 1}_c\over \sqrt{N_c}}\right), \label{hc:projector2} \eqa
where $N_c=3$, and ${\bf 1}_c$ stands for the unit color matrix. In
above, $M_{\eta_b}=2m_b$ and $M_{J/\psi}=2m_c$ are implicitly
assumed.

\begin{figure}
\centering
\includegraphics[width=1\textwidth]{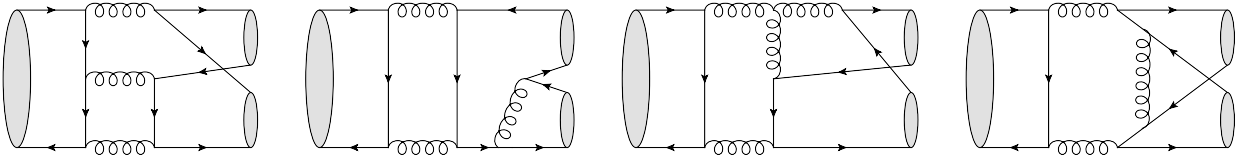}%
\hspace*{\fill} \caption{\small Typical Feynman diagrams of the
exclusive process $\eta_b(P_{\eta_b}) \rightarrow
J/\psi(P_{J/\psi_1}) + J/\psi(P_{J/\psi_2})$ at the one-loop level.}
\label{lpty} \vspace{-0mm}
\end{figure}

For this process, at the leading order of relative velocity $v$ in
the framework of NRQCD, the tree level feynman-diagram has no
contribution to the branching ratio, since the trace of the b-quark
line form a Lorentz antisymmetric tensor, while the trace of c-quark
line form a Lorentz symmetric tenser. This situation remains also in
the NLO counterterm, self-energy and vertex correction diagrams.
Therefore, at one-look level only a few types of Feynman diagrams
should be taken into account in the calculation, which are
schemetically showed in Figure \ref{lpty}.

Because of parity and Lorentz invariance, the decay amplitude
possesses the following unique tensor structure:
\bqa
\mathcal{M}(\lambda_1,\lambda_2)=\mathcal{A}\;
\varepsilon_{\mu\nu\rho\sigma}\varepsilon_{J/\psi_1}^{*\mu}
(\lambda_1)
\varepsilon_{J/\psi_2}^{*\nu}(\lambda_2)
P^\rho_{J/\psi_1}P^\sigma_{J/\psi_1}\,.
\label{hc:projector} \eqa
In our calculation, the computer algebra system MATHEMATICA is
employed with the help of the packages, FYENCALC \cite{Feyncalc},
FYENART \cite{Feynart}, and LoopTools \cite{Looptools}. FYENART is
used to draw Feynman diagrams and generate amplitudes, FYENCALC is
used to trace the $\gamma$ matrices and to reduce various
Passarino-Veltman tensor integrals \cite{PaVe} to scalar ones,
LoopTools is used to evaluate these integrals. For the aim of
comparison, we express all the Passarino-Veltman scalar integrals
encountered in this calculation in the Appendix.

After taking the above mentioned procedures, it is straightforward
to calculate this process and obtain the analytical amplitude in
reduced form. i.e.,
\bqa
\mathcal{A}=\frac{512\sqrt{2}\pi\alpha_s^3m_c
\psi_{\eta_b}(0)\psi^2_{J/\psi}(0)}
{9\sqrt{3}m_b^{9/2}(m_b^2-4m_c^2)}\; F\left(m_c^2,m_b^2\right)\, ,
 \label{hc:projector} \eqa
where
\bqa
F\left(m_c^2,m_b^2\right)=&&-\frac{1}{4} \text{D}_0(2) m_b^4+2
\text{D}_0(1) m_c^2 m_b^2+\text{D}_0(2) m_c^2 m_b^2+\frac{19}{16}
\text{C}_0(1) m_b^2+\frac{3}{2} \text{C}_0(2) m_b^2\nonumber\\
&&-\frac{9}{4}
   \text{C}_0(3) m_b^2-\frac{1}{4} \text{C}_0(5)
   m_b^2-\frac{1}{4} \text{C}_0(6) m_b^2
   -\frac{9}{16} \text{C}_0(7) m_b^2-\frac{9}{4}
   \text{C}_0(1) m_c^2\nonumber\\
&&-\frac{7}{4}
   \text{C}_0(2) m_c^2-2 \text{C}_0(4) m_c^2+\frac{9
   \text{B}_0(1)}{8}
   -\frac{9 \text{B}_0(2)}{4}+2 \text{B}_0(3)-\frac{7
   \text{B}_0(4)}{8}\nonumber\\
&&+\frac{9 \text{B}_0(5)}{8}-\frac{9
   \text{B}_0(6)}{8}+\frac{\text{B}_0(3) m_c^2}{m_b^2}-
   \frac{\text{B}_0(6) m_c^2}{m_b^2}
 \label{form}\,. \eqa
Here, the form factor $F\left(m_c^2,m_b^2\right)$ is a complex
function; $\text{D}_0$, $\text{C}_0$, and $\text{B}_0$ represent
four-point, three-point and two-point Passarino-Veltman scalar
integrals, respectively. The real part of
$F\left(m_c^2,m_b^2\right)$ is too complicated to be presented here,
and therefore only the asymptotic form in small $m_c$ limit is
given:
\bqa
\textmd{Re}(F\left(m_c^2,m_b^2\right))_{asy}=&&\frac{19}{32}
\textmd{log}^2(a)-
\frac{1}{8}\textmd{log}(2)\textmd{log}(a)
+\frac{5}{4}\textmd{log}(a)
+\frac{5}{16}\textmd{log}^2(2)\nonumber\\
&&+\frac{1}{2}\textmd{log}(2)+\frac{29\pi^2}{96}-\frac{3
\sqrt{3}}{8}\pi+\frac{3}{4}
 \label{hc:projector} \eqa
with $a=\frac{m^2_c}{m^2_b}$. The full imaginary part of
$F\left(m_c^2,m_b^2\right)$ is
\bqa
\textmd{Im}(F\left(m_c^2,m_b^2\right))=&&\frac{(36a-19)
\pi}{16\delta}\textmd{log}(\frac{1 +\delta}{1-\delta})
-\frac{(36a-5)\pi}{16\delta}
\textmd{log}(\frac{3-\delta}{3+\delta})\nonumber\\&&
+\frac{\pi}{4\delta}\textmd{log}((1 +\delta)^2(1+\delta^2))
+\frac{(2\delta^2+7\delta+7)\pi}{8(\delta+1)} \,,
\label{hc:projector} \eqa
where $\delta=\sqrt{1-4a}$, and its asymptotic form in the small
$m_c$ limit reads
\bqa
\textmd{Im}(F\left(m_c^2,m_b^2\right))_{asy}=
\frac{19\pi}{16}\textmd{log}(a)
+\frac{7\pi}{16}\textmd{log}(2)+\pi\,.
 \label{hc:projector} \eqa
With the above preparation, we can readily obtain the branching
fraction of the exclusive $\eta_b \rightarrow J/\psi J/\psi$ decay
process,

\bqa
Br[\eta_b \rightarrow J/\psi
J/\psi]=K_{gg}^{-1}\frac{2^{13}\alpha_s^4m_c^2
\psi^4_{J/\psi}(0)}{3^4m_b^7\sqrt{m_b^2
-4m_c^2} }|F(m_c^2,m_b^2)|^2 \,.\eqa
Here, the dominant $\eta_b$ gluonic decay width is taken to be its
total width approximately at one-loop order \cite{NLOwidth}, i.e.,
\bqa
\Gamma[\eta_b]_{total}\approx\Gamma_{NLO}[\eta_b\rightarrow g
g]=K_{gg}\frac{8\pi\alpha_s^2}{3m_b^2}\psi^2_{\eta_b}(0) \eqa
with
\bqa
K_{gg}=1+(C_F(-5+\frac{\pi^2}{4})+C_A(\frac{199}{18}-\frac{13\pi^2}{24})
-\frac{16}{9}n_f T_F)\frac{\alpha_s(2m_b)}{\pi}\; .\eqa

In numerical calculation, the following inputs are adopted:
\bqa
\psi_{J/\psi}(0)=0.263\ \textmd{GeV}^{3/2},\; m_c=1.5\ \textmd{GeV},
\; m_b=4.7\ \textmd{GeV}, \; \alpha_s=0.18\sim0.26\, , \eqa
where the radial wave function at the origin $\psi_{J/\psi}(0)$ is
obtained by fitting the NLO QCD calculation result to the $J/\psi$
di-lepton decay width \cite{pdg}. With the above preparation, we can
readily obtain the numerical result of the concerned process
\bqa
Br[\eta_b \rightarrow J/\psi
J/\psi]=5.93\times10^{-8}\sim2.58\times10^{-7}\,. \eqa
Here, the uncertainties are originated from energy scale variation
from $m_b$ to $m_c$. It is worth emphasizing that in our result the
double logarithms exist and agree with what obtained in Ref.
\cite{b13}, whereas our constant term does not agree with theirs,
though its numerical influence is no big.

\section{Higher twist contributions}

As mentioned in preceding sections, at Born level $\Gamma
(\eta_b\rightarrow J/\psi J/\psi )$ is exactly zero in
non-relativistic limit, while the NLO radiative corrections are very
small. People find that although in light cone formalism the leading
twist term in the light cone distribution amplitudes(LCDAs) for
$J/\psi$ vanishes in $\eta_b\rightarrow J/\psi J/\psi$ process,
contributions from higher twist terms seem to be important
\cite{b14}. In Ref. \cite{b14} the LCDAs up to twist-4 are taken
into account for the consistency reason. It is true and in the
following we reevaluate this process also in the light cone
framework. However, to execute the twist expansion accurately, for
final vector mesons with transverse polarizations, we expand the
LCDA projector in momentum space given by \cite{beneke} to twist-4,
which yields more terms than what employed in Ref. \cite{b14}. i.e.,
\begin{eqnarray}
M^V_\perp&=&(M^{(2)}_\perp+M^{(3)}_\perp+M^{(4)}_\perp)\Big|_{k=up}\;
,
\end{eqnarray}
where
\begin{eqnarray}
M^{(2)}_\perp&=&\frac{1}{4}f^T_V E\slashchar{\varepsilon}_\perp
\slashchar{n}_-\phi_\perp(u)\;,\\
M^{(3)}_\perp&=&\frac{1}{4}f_Vm_V\left[\slashchar{\varepsilon}_\perp
g_\perp^{(v)}(u)-E\slashchar{n}_-
\int^u_0\!dv\left(\phi_\parallel(v)-g_\perp^{(v)}(v)\right)
\varepsilon_\perp^\sigma\frac{\partial}{\partial
k_\perp^\sigma}\right]\nonumber\\
&+&\frac{i}{4}\left(f_{V} - f_{V}^{T} \frac{m_{1} +
m_{2}}{m_{V}}\right) m_{V} \varepsilon_{\mu\nu\rho \sigma}
\varepsilon_\perp^{\nu} n_-^{\rho}\gamma^\mu\gamma_5
\left(n_+^\sigma\frac{g'^{(a)}_\perp(u)}{8}-E\frac{
g^{(a)}_\perp(u)}{4}\frac{\partial}{\partial
k_{\perp\sigma}}\right)\;, \\
M^{(4)}_\perp&=&\frac{1}{4}\frac{m^2_V}{E}f_V^T\Bigg[\frac{1}{4}
\slashchar{\varepsilon}_\perp\slashchar{n}_+
h_3(u)-\frac{E}{4}[\slashchar{n}_-,\slashchar{n}_+] \int_0^u\! dv
\left(h_\parallel^{(t)}(v)-\frac{1}{2}
\phi_\perp(v)-\frac{1}{2}h_3(v)\right)
\varepsilon_\perp^\sigma\frac{\partial}{\partial
k_\perp^\sigma}\Bigg]\nonumber\\
&&-\frac{1}{4}m^2_V\left(f_{V}^T - f_{V} \frac{m_{1} +
m_{2}}{m_{V}}\right)
\frac{h^{(s)}_\parallel(u)}{2}
\varepsilon_\perp^\sigma\frac{\partial}{\partial
k_\perp^\sigma}\;,
\end{eqnarray}
with $E=(p^0+|\vec{p}|)/2$ and the transverse polarization vector
\begin{eqnarray}
\epsilon_\perp^\mu=\varepsilon^\mu-\frac{\varepsilon\cdot
n_+}{2}n_-^\mu-\frac{\varepsilon\cdot n_-}{2}n_+^\mu\; .
\end{eqnarray}
The higher twist LCDAs  are related to the twist-2 ones by the
Wandzura-Wilczek relations\cite{qsr1},
\begin{eqnarray}
g^{(v)}_\perp(u)&=&\frac{1}{2}\left[\int^u_0\frac{\phi_\parallel(v)}
{\bar{v}}dv+\int^1_u\frac{\phi_\parallel(v)}{v}dv\right],\label{ww1}\\
g^{(a)}_\perp(u)&=&2\left[\bar{u}\int^u_0\frac{\phi_\parallel(v)}
{\bar{v}}dv+u\int^1_u\frac{\phi_\parallel(v)}{v}dv\right],\label{ww2}\\
h^{(s)}_\parallel(u)&=&2\left[\bar{u}\int^u_0\frac{\phi_\perp(v)}
{\bar{v}}dv+u\int^1_u\frac{\phi_\perp(v)}{v}dv\right]\label{ww3}.
\end{eqnarray}
Here, the contributions from three-particle DAs have been neglected
as performed in Ref. \cite{b14}.

After a lengthy calculation, we get the final expression for the
decay amplitude of the process $\eta_Q\rightarrow V_1V_2$,
\begin{eqnarray}\label{analytical_result}
\mathcal{M}_{\perp\perp}&=&T_0\varepsilon_{\mu\nu\rho\sigma}
\varepsilon_{1\perp}^{*\mu}\varepsilon_{2\perp}^{*\nu}
n_-^\rho
n_+^\sigma\int^1_0\!du_1\int^1_0\!d\bar{u}_2\frac{1}{256E_1^2E^2_2
u_1\bar{u}_1u_2\bar{u}_2(u_1u_2+\bar{u}_1\bar{u}_2)}\nonumber\\
&\times&\Bigg\{m_{V_1}m_{V_2}f_{V_1}\tilde{f}_{V_2}
\Big[g_{1\perp}^{(v)}(u_1)g^{(a)}_{2\perp}(\bar{u}_2)+
\Phi_1(u_1)g^{'(a)}_{2\perp}(\bar{u}_2)+f(u_1,\bar{u}_2)
\Phi_1(u_1)g^{(a)}_{2\perp}
(\bar{u}_2)\Big]\nonumber\\
&&+m_{V_1}m_{V_2}f_{V_2}\tilde{f}_{V_1}\Big[g_{2\perp}^{(v)}
(\bar{u}_2)g^{(a)}_{1\perp}(u_1)+
\Phi_2(\bar{u}_2)g^{'(a)}_{1\perp}(u_1)-f(u_1,\bar{u}_2)
\Phi_2(\bar{u}_2)g^{(a)}_{1\perp}(u_1)\Big]\nonumber\\
&&+2m^2_{V_2}
f^T_{V_1}\tilde{f}^T_{V_2}\phi_{1\perp}
(u_1)h^{(s)}_{2\parallel}(\bar{u}_2)+
2m^2_{V_1}f^T_{V_2}\tilde{f}^T_{V_1}\phi_{2\perp}
(\bar{u}_2)h^{(s)}_{1\parallel}(u_1)\Bigg\}\;,
\end{eqnarray}
with
\begin{eqnarray}
&&T_0=8g_s^4\frac{\psi_{\eta_Q}(0)}{\sqrt{8m_Q}}\frac{N_c^2-1}
{4N_c^2\sqrt{N_c}}\;,\\
&&f(u_1,\bar{u}_2)=(u_1-\bar{u}_2)\left(\frac{-1}{u_1\bar{u}_2}+
\frac{-1}{u_2\bar{u}_1}+\frac{2}{u_1u_2+\bar{u}_1\bar{u}_2}\right),\\
&&\tilde{f}_{V}=f_{V} - f_{V}^{T} \frac{m_{1} + m_{2}}{m_{V}}\;,
\quad\tilde{f}^T_{V}=f_{V}^T - f_{V} \frac{m_{1} + m_{2}}{m_{V}}\;,\\
&&\Phi_1(u_1)=\int^{u_1}_0\!dw
\Big(\phi_{1\parallel}(w)-g_{1\perp}^{(v)}(w)\Big)\;,\quad
\Phi_2(\bar{u}_2)=\int^{\bar{u}_2}_0\!dw
\Big(\phi_{2\parallel}(w)-g_{2\perp}^{(v)}(w)\Big)\;.
\end{eqnarray}
Here, $\psi_{\eta_Q}(0)$ is the wave function at the origin for
pseudoscalar $\eta_Q$. Our analytical result is different from what
given in Ref.\cite{b14}, partly due to the different projectors
used.

With the asymptotic form for twist-2 distribution amplitudes,
\begin{eqnarray}
\phi_\perp(u)=\phi_\parallel(u)=\phi_{AS}(u)=6u(1-u)\;,
\end{eqnarray}
the analytical decay amplitude turns to be pretty simple, it reads
\begin{eqnarray}
\mathcal{M}_{\perp\perp}&=&T_0\varepsilon_{\mu\nu\rho\sigma}
{\varepsilon_{1\perp}^{*\mu}\varepsilon_{2\perp}^{*\nu} n_-^\rho
n_+^\sigma}\frac{9}{256E^2_1E^2_2}
\times\Big[(\pi^2-4)m_{V_1}m_{V_2}(f_{V_1} \tilde{f}_{V_2} +
f_{V_1}\tilde{f}_{V_2})\nonumber\\ && +\ 2\pi^2(m^2_{V_2}
f^T_{V_1}\tilde{f}^T_{V_2}+m^2_{V_1}
f^T_{V_2}\tilde{f}^T_{V_1})\Big]\;.
\end{eqnarray}

To numerically evaluate the branching ratio of $\eta_b\rightarrow
J/\psi J/\psi$ process, we use the following commonly accepted input
parameters: the charm quark mass in the $\overline{MS}$ scheme,
$m_c^{\overline{MS}}=1.2$ GeV; the $J/\psi$ decay constant
$f_{J/\psi}=416$ MeV; and the $f^T_{J/\psi}$ is obtained in the
framework of NRQCD \cite{ftjpsi}, $f^T_{J/\psi}=379$ MeV. Then the
numerical result reads
\begin{eqnarray}
Br[\eta_b\rightarrow J/\psi J/\psi]=(1.1\sim 2.3)\times10^{-6}\; .
\end{eqnarray}
Here, the uncertainties are also induced by the scale variation from
$m_b$ to $m_c$ as in above NLO QCD calculation. Note that the above
magnitude is bigger than what the NLO result. This is however
understandable, since roughly speaking the higher twist
contributions are suppressed by factor of
$\Big(\frac{m_c}{m_b}\Big)^4$ while the NLO contributions are
suppressed by $\alpha_s^2\Big(\frac{m_c}{m_b}\Big)^2$.

Following we apply the above higher twist analysis to the $\eta_c$
to light vector mesons decay process for the first time to twist-4.
As mentioned in the introduction the disagreement of experimental
measurement with theoretical expectation is a long lasting issue.
Before attributing some non-perturbative scenarios, to evaluate
these processes in light cone formalism to twist-4 is still
meaningful. We already know that the leading twist term in LCDAs at
leading order of $\alpha_s$ does not contribute to these processes,
and the radiative corrections are dramatically suppressed by the
light quark masses. Therefore, it is obvious that contributions from
higher twist DAs dominate over others in the framework of
perturbative QCD. In addition to the asymptotic form, the LCDA form
in terms of Gegenbauer polynomials is also employed. That is
\begin{eqnarray}
\phi_{\parallel,\perp}(u,\mu^2)=
6u(1-u)\left(1+\sum^\infty_{n=1}a^{\parallel,\perp}_n(\mu^2)
C_n^{3/2}(2u-1)\right)\; .
\end{eqnarray}
The input parameters needed in the numerical calculation are listed
in TABLE. \ref{parameter}, $f^T_V$s come from the QCD sum rules
\cite{qsr1,qsr2,qsr3,qsr4}, and the reasonable values of Gegenbauer
moments $a_1$ and $a_2$ are from Ref. \cite{ftvalue}. All
scale-dependent quantities refer to $\mu=1$ GeV.
\begin{table}[h]\caption{Summary of theoretical input parameters.}
\label{parameter}
\begin{tabular}{ccccccccc}\hline
&$\hspace{1.5cm}\rho\hspace{1.5cm}$&$\hspace{1.5cm}\bar{K}^*
\hspace{1.5cm}$&$\hspace{1.5cm}\omega
\hspace{1.5cm}$&$\hspace{1.5cm}\phi\hspace{1.5cm}$\\
\hline $m_V$[{\rm MeV}]&770&892&782&1020 \\ $f_V[{\rm MeV}]$
&$205\pm
9$&$217\pm5$&$195\pm3$&$231\pm4$\\
$f^T_V[{\rm MeV}]$ &$160\pm10$&$170\pm10$&$145\pm10$&$200\pm10$\\
$a_1^{\parallel,\perp}$ &0&$0.10\pm0.07$&0&0\\
$a_2^{\parallel,\perp}$
&$0.09^{+0.10}_{-0.07}$&$0.07^{+0.09}_{-0.07}$
&$0.09^{+0.10}_{-0.07}$&$0.06^{+0.09}_{-0.07}$\\
\hline
\end{tabular}
\end{table}
The numerical results are given in Table. \ref{result}, where Br[AS]
and Br[GP] represent the results for forms of asymptotic and the
Gegenbauer polynomials in LCDAs, respectively. Since the Gegenbauer
moments $a_n$ are small, the numerical results are not sensitive to
the form of the leading distribution amplitudes, which influence the
higher twist results via relations (\ref{ww1})-(\ref{ww3}). From
results in Table \ref{result} we see that although the higher twist
effect are tremendous for Br$[\eta_c\rightarrow VV]$, it is still
not enough to explain the experimental data.

\begin{table}[h]\caption{Experimental data and Numerical results for
$Br[\eta_c\rightarrow VV]$, experimental data are from Particle Data
Book \cite{pdg}.}\label{result}
\begin{tabular}{cccc}\hline
Final state&$\hspace{1.5cm}{\rm Br[ex]}\hspace{1.5cm}$&$
\hspace{1.5cm}{\rm Br[AS]}
\hspace{1.5cm}$&$\hspace{1.5cm}{\rm Br[GP]}\hspace{1.5cm}$\\
\hline
$\rho\rho$&$(2.0\pm0.7)\times 10^{-2}$ &$2.0\times10^{-4}$
&$2.8\times10^{-4}$\\
$K^*\bar{K}^*$ &$(9.2\pm3.4)\times 10^{-3}$ &$7.2
\times10^{-4}$ &$9.0\times10^{-4}$\\
$\omega\omega$ &$<3.1\times 10^{-3}$ & $9.1
\times10^{-5}$&$1.3\times10^{-4}$ \\
$\phi\phi$&$(2.7\pm0.9)\times 10^{-3}$ & $6.6
\times10^{-4}$&$8.1\times10^{-4}$\\
\hline
\end{tabular}
\end{table}

\section{Conclusions}

To further study the nature of recently observed state $\eta_b$,
direct measurement of its decay products is necessary. The
$\eta_b\rightarrow J/\psi J/\psi$ process was considered and
suggested to be a golden channel to this aim. In the literature,
different theoretical estimation varies greatly. The branching ratio
starts from $10^{-4}$ to $10^{-8}$, which induces some confusion for
future experimental test. Within the pQCD and factorization scheme
we have calculated this helicity conservation suppressed process,
the $\eta_b\rightarrow J/\psi J/\psi$, at the next-to-leading order
in pQCD. Our result confirms the existence of double logarithms, the
$\textmd{log}^2(\frac{m_c^2}{m_b^2})$ in Ref. \cite{b13}, and the
coefficients of both double logarithm
$\textmd{log}^2(\frac{m_c^2}{m_b^2})$ and single logarithm
$\textmd{log}(\frac{m_c^2}{m_b^2})$ in our calculation are
consistent with those in the same reference.  However, we find that
other terms in our result deviate from those in Ref. \cite{b13},
though the numerical significance of the difference is not high.

In the light cone formalism, the leading twist contribution to
$\eta_b\rightarrow J/\psi J/\psi$ process vanishes. In this work we
also evaluate the higher twist contributions to it. Expanding the
LCDAs of final vector mesons to twist-4, we find that the higher
twist terms contribute more to the decay width than what from the
NLO corrections, which implies that the final state mass effects is
more significant than the NLO corrections in this helicity
suppressed process. According to our twist-4 calculation, the
branching fraction of $\eta_b\rightarrow J/\psi J/\psi$ process can
be as large as $\sim 10^{-6}$, which enables the direct search of
$\eta_b$ in Tevatron Run II or LHC feasible. In Ref. \cite{b14}, the
same process was evaluated in the light cone formalism also to
twist-4, but with different twist expansion procedure, which lead to
different LCDAs from ours. We believe what we used are generated
from the LCDA definition in twist expansion and should be more
proper.

Unlike the undiscovered $\eta_b\rightarrow J/\psi J/\psi$ process,
experimental results about $\eta_c\rightarrow VV$ indicate that
relatively large violations of the helicity conservation rules exist
in these processes. The surprisingly large branching ratios of
$\eta_c\rightarrow VV$ still stand as a bewildering puzzle in
charmonium physics. Since it is believed that the NLO corrections
are greatly suppressed by the light quark mass, the higher twist
effects might be large. In the light cone formalism, we have
calculated this process with taking the next-to-next-leading twist
effects in LCDAs of final vector mesons into account. Result shows
that the higher twist DAs indeed violate the helicity conservation
rules, though it still deviates a lot from the experimental
measurement. This implies that the perturbative description of
$\eta_c$ decay alone is not enough, and some non-perturbative
mechanism may play important roles in $\eta_c$ decays, such as
$\eta_c\rightarrow VV$, which deserves further investigation
\cite{haoqiao}.

\vspace{1cm}

{\bf Acknowledgments}

This work was supported in part by the National Natural Science
Foundation of China(NSFC) under the grants 10935012, 10928510,
10821063 and 10775179, by the CAS Key Projects KJCX2-yw-N29 and
H92A0200S2.

\vspace{1cm}

{\bf Appendix} \vspace{.3cm}

In this appendix, we list various Passarino-Veltman scalar integrals
appearing in Eq.(\ref{form}), and only the leading power in $a$ for
real part are extracted; while for imaginary part, we present the
full expression. Since coefficients of the $\textmd{D}_0(1)$ and
$\textmd{C}_0(4)/m_b^2$ in Eq.(\ref{form}) are same up to a sign,
hence we only present their difference. Our evaluation for these
integrals agrees with LoopTools, and we have also checked with
FIESTA2 \cite{AV} the asymptotic expression. Here,
$a=\frac{m_c^2}{m_b^2}$ and $\delta=\sqrt{1-4a}$.
\bqa
\textmd{D}_0(1)&&=\textmd{D}_0[m_b^2, m_c^2, 4m_c^2, m_c^2, m_c^2,
2m_b^2 + m_c^2, 0, 0,
m_c^2, m_c^2]\nonumber\\
&&=\frac{\textmd{C}_0(4)}{m_b^2}-\frac{\textmd{log}(2)}
{m_c^2m_b^2}\\\nonumber
\\
\textmd{D}_0(2)&&=\textmd{D}_0[m_b^2, m_c^2, 4m_b^2, m_c^2, 2m_b^2 +
m_c^2, 2m_b^2
 + m_c^2, m_c^2, m_c^2, 0, 0]\nonumber\\
&&\approx\frac{1-\textmd{log}(a)}{2m_b^4}-
\mathbbm{i}\pi\frac{1}{4m_b^2(1+\delta)}\\\nonumber
\\
\textmd{C}_0(1)&&=\textmd{C}_0[m_b^2, m_c^2,
m_c^2, 0, 0, m_c^2]\nonumber\\
&&\approx\frac{3\textmd{log}^2(a)+\pi^2}{6m_b^2}-
\mathbbm{i}\pi\frac{\textmd{log}(\frac{1+\delta}{1-\delta})}{
m_b^2\delta}\\\nonumber
\\
\textmd{C}_0(2)&&=\textmd{C}_0[m_b^2, m_c^2,
2m_b^2 + m_c^2, 0, 0, m_c^2]\nonumber\\
&&\approx\frac{-6\textmd{log}(2)\textmd{log}
(a)+3\textmd{log}^2(2)+\pi^2}{6m_b^2}-
\mathbbm{i}\pi\frac{\textmd{log}(\frac{3-\delta}{3+\delta})}{
m_b^2\delta}\\\nonumber
\\
\textmd{C}_0(3)&&=\textmd{C}_0[4m_b^2, m_c^2,
2m_b^2 + m_c^2, 0, 0, m_c^2]\nonumber\\
&&\approx\frac{-6\textmd{log}(2)\textmd{log}(a)+
3\textmd{log}^2(2)+\pi^2}{12m_b^2}-
\mathbbm{i}\pi\frac{\textmd{log}(\frac{3-\delta}{3+\delta})}{
2m_b^2\delta}\\\nonumber
\\
\textmd{C}_0(4)&&=\textmd{C}_0[m_c^2, m_c^2, 4m_c^2, m_c^2, 0,
m_c^2]\\\nonumber
\\
\textmd{C}_0(5)&&=\textmd{C}_0[m_b^2, m_c^2,
m_c^2, m_c^2, m_c^2, m_b^2]\nonumber\\
&&\approx-\frac{\pi^2}{12m_b^2}-
\mathbbm{i}\pi\frac{\textmd{log}(2-4a)}{ m_b^2\delta}\\\nonumber
\\
\textmd{C}_0(6)&&=\textmd{C}_0[m_b^2, m_c^2,
2m_b^2 + m_c^2, m_c^2, m_c^2, 0]\nonumber\\
&&\approx\frac{-6\textmd{log}(2)\textmd{log}(a)-
3\textmd{log}^2(2)+\pi^2}{6m_b^2}-
\mathbbm{i}\pi\frac{\textmd{log}(\frac{(3-
\delta)(1+\delta)^2}{3+\delta})}{
m_b^2\delta}\\\nonumber
\\
\textmd{C}_0(7)&&=\textmd{C}_0[m_b^2, m_c^2, m_c^2, m_b^2, m_b^2,
m_c^2]\approx-\frac{\pi^2}{9m_b^2}\\\nonumber
\\
\textmd{B}_0(1)&&=\textmd{B}_0[m_b^2, 0,
0]\approx\frac{1}{\epsilon}+2-\textmd{log}(m_b^2)+
\mathbbm{i}\pi\\\nonumber
\\
\textmd{B}_0(2)&&=\textmd{B}_0[4m_b^2, 0,
0]=\frac{1}{\epsilon}+2-\textmd{log}(4m_b^2)+
\mathbbm{i}\pi\\\nonumber
\\
\textmd{B}_0(3)&&=\textmd{B}_0[2m_b^2 + m_c^2, 0,
m_c^2]\approx\frac{1}{\epsilon}+2-\textmd{log}(2m_b^2)+
\mathbbm{i}\pi\frac{2}{2+a}\\\nonumber
\\
\textmd{B}_0(4)&&=\textmd{B}_0[m_c^2, m_b^2,
m_c^2]\approx\frac{1}{\epsilon}+1-\textmd{log}(m_b^2)\\\nonumber
\\
\textmd{B}_0(5)&&=\textmd{B}_0[m_b^2, m_b^2,
m_b^2]=\frac{1}{\epsilon}+2-\frac{\pi}{\sqrt{3}}-
\textmd{log}(m_b^2)\\\nonumber
\\
\textmd{B}_0(6)&&=\textmd{B}_0[m_c^2, 0,
m_c^2]=\frac{1}{\epsilon}+2-\textmd{log}(m_c^2)\\\nonumber
 \eqa

\end{document}